\documentclass[twocolumn,aps,pra,superscriptaddress]{revtex4-1}
\usepackage[utf8]{inputenc}
\usepackage[T1]{fontenc}
\usepackage[english]{babel}
\usepackage{bm}
\usepackage{physics}
\usepackage{amsfonts}
\usepackage[normalem]{ulem} 
\usepackage{tcolorbox}
\usepackage{wrapfig}

\usepackage{hyperref}

\hypersetup{
colorlinks=true,
citecolor=blue,
linkcolor=red,
urlcolor=blue
 ,
 pdfmenubar=true
}

\tcbuselibrary{breakable}

\newtcolorbox[use counter=myexample,number format=\Alph 
]{mybox}[2][]{%
/tcb/breakable,/tcb/bottomrule
at break=-1pt,/tcb/toprule
at break=-1pt,colback=black!5!white,colframe=black!80!white,fonttitle=\bfseries,
title=Box \thetcbcounter: #2,#1}

\begin{document}
\title{Learning Quantum Systems}

\author{Valentin Gebhart}
\affiliation{QSTAR,  INO-CNR  and  LENS,  Largo  Enrico  Fermi  2,  50125  Firenze,  Italy}
\author{Raffaele Santagati}
\affiliation{Boehringer Ingelheim, Quantum Lab, Doktor-Boehringer-Gasse 5-11, 1120 Vienna, Austria}
\author{Antonio Andrea Gentile}
\affiliation{Pasqal SAS, 7 Rue L. de Vinci, 91300, Massy, France}
\author{Erik M. Gauger}
\affiliation{School of Engineering and Physical Sciences, SUPA, Heriot-Watt University, Edinburgh, EH14 4AS, United Kingdom}
\author{David Craig}
\affiliation{Department of Materials, University of Oxford, Oxford OX1 3PH, United Kingdom}
\author{Natalia Ares}
\affiliation{Department of Engineering Science, University of Oxford, Oxford OX1 3PJ, United Kingdom}
\author{Leonardo Banchi}
\affiliation{Department of Physics and Astronomy, University of Florence,
via G. Sansone 1, I-50019 Sesto Fiorentino (FI), Italy}
\affiliation{ INFN Sezione di Firenze, via G. Sansone 1, I-50019, Sesto Fiorentino (FI), Italy }
\author{Florian Marquardt}
\affiliation{Max Planck Institute for the Science of Light and Friedrich-Alexander-Universit\"at Erlangen-N\"urnberg, Erlangen, Germany}
\author{Luca Pezzè}
\email{luca.pezze@ino.cnr.it}
\affiliation{QSTAR,  INO-CNR  and  LENS,  Largo  Enrico  Fermi  2,  50125  Firenze,  Italy}
\author{Cristian Bonato}
\email{c.bonato@hw.ac.uk}
\affiliation{School of Engineering and Physical Sciences, SUPA, Heriot-Watt University, Edinburgh, EH14 4AS, United Kingdom}

\begin{abstract}
The future development of quantum technologies relies on creating and manipulating quantum systems of increasing complexity, with key applications in computation, simulation and sensing. This poses severe challenges in the efficient control, calibration and validation of quantum states and their dynamics. Although the full simulation of large-scale quantum systems may only be possible on a quantum computer, classical characterization and optimization methods still play an important role. Here, we review different approaches that use classical post-processing techniques, possibly combined with adaptive optimization, to learn quantum systems, their correlation properties, dynamics and interaction with the environment. We discuss theoretical proposals and successful implementations across different multiple-qubit architectures such as spin qubits, trapped ions, photonic and atomic systems, and superconducting circuits. This Review provides a brief background of key concepts recurring across many of these approaches with special emphasis on the Bayesian formalism and neural networks. 
\end{abstract}

\maketitle


\section*{ Introduction}

The development of efficient techniques for the characterization of quantum systems is motivated by the potential impact of quantum technologies in communication, computing, sensing and simulation. 
However, the complexity of quantum states and dynamics increases exponentially with their size, making their full description intractable and even approximations challenging. 
 
Learning quantum systems (that is, acquiring essential information about them) is crucial for quantum technologies regardless of the particular application or physical architecture. 
For instance, in the context of quantum computation, it is important to calibrate and benchmark the quantum gates and operations implemented in the actual device, and to prepare qubit states with high fidelity.
In quantum metrology, the optimization of probe states, parameter encoding transformations and measurement strategies is pivotal for improving the accuracy and precision of the sensor. 
In quantum simulation, although the full simulation of large quantum systems may only be possible with a quantum computer \cite{Feynman1982}, classical methods for the characterization of systems of medium size \cite{Preskill2018} (of the order of hundreds of qubits~\cite{AruteNATURE2019, ZhongSCIENCE2020, EbadiNATURE2021}) are currently playing a key role.
Such approaches have been successfully applied to experiments, supporting the exploration of fundamental properties such as entanglement, quantum superpositions and nonlocality.

\begin{figure*}[ht]
\centering
\includegraphics[width=.95\linewidth]{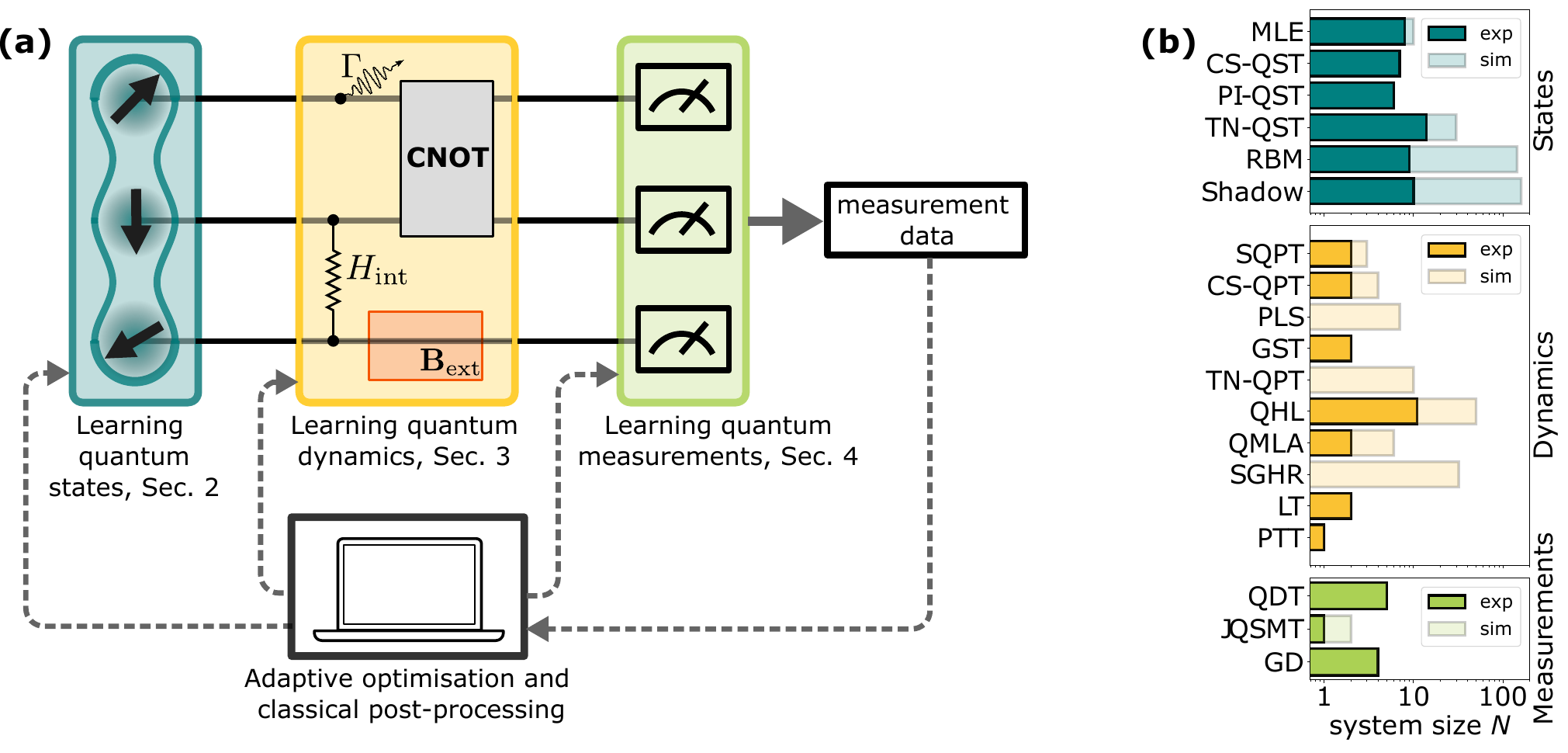}
\caption{\textbf{Learning quantum states, dynamics and measurements.}
\textbf{a} In this Review, we divide the task of learning quantum systems into the sub-tasks of learning quantum states,  quantum dynamics and  quantum measurements. 
Examples of quantum processes are decoherence (schematically represented as $\Gamma$ in the figure), particle interaction ($H_{\rm int}$), quantum gates (controlled NOT gate, $\operatorname{CNOT}$) and coupling to an external field ($\mathbf{B}_{\rm ext}$).
\textbf{b} A list of widely used methods for learning quantum states (top), dynamics (centre), and measurements (bottom), illustrating the number of qubits $N$ each method has been applied to, for experimental (exp) and simulated (sim) data. The compared methods are maximum-likelihood estimation (MLE)~\cite{haffner_scalable_2005, shang_superfast_2017, bolduc_projected_2017}, compressed-sensing quantum-state tomography (CS-QST)~\cite{riofrio_experimental_2017}, permutationally-invariant QST (PI-QST)~\cite{schwemmer_experimental_2014}, tensor-network QST (TN-QST)~\cite{cramer_efficient_2010,lanyon_efficient_2017}, restricted Boltzmann machines (RBMs)~\cite{torlai_neural-network_2018,torlai_integrating_2019}, and classical shadows (shadow)~\cite{elben_mixed-state_2020,huang_predicting_2020} for quantum states; standard quantum-process tomography (SQPT)~\cite{Altepeter_2003,Rodionov_2014,knee_2018,torlai_quantum_2020}, compressed-sensing QPT (CS-QPT)~\cite{Shabani_2011,Rodionov_2014}, projected-least-squares (PLS)~\cite{surawy2021projected}, gate set tomography (GST)~\cite{nielsen_gate_2020,zhang2020error,song2019quantum}, tensor-network QPT (TN-QPT)~\cite{torlai_quantum_2020}, quantum Hamiltonian learning (QHL)~\cite{hou_2019_11qubit,wiebe_bootstrapping_2015}, quantum model learning agent (QMLA)~\cite{gentile_learning_2021}, state-guided Hamiltonian reconstruction (SGHR)~\cite{zhu2019reconstructing,anshu_sample-efficient_2021}, Lindblad tomography (LT)~\cite{samach2022lindblad}, and process tensor tomography (PTT)~\cite{white2022-PTT} for quantum dynamics; and quantum detector tomography (QDT)~\cite{ChenPRA2019}, joint quantum-state and measurement tomography (JQSMT)~\cite{ZhangPRL2020,KeithPRA2018}, and gradient descend methods (GD)~\cite{hetzel_tomography_2022} for quantum measurements.
Note that some of the above methods, even though not fully machine-learning-inspired, still rely on machine-learning techniques to perform key subroutines, as detailed in the main text.}
\label{fig:fig1_summary}
\end{figure*}

Here we provide a broad overview of efficient learning techniques that use classical post-processing and adaptive optimization to learn quantum states, quantum dynamics and quantum measurements (see Fig.~\ref{fig:fig1_summary}).
Different methods may provide full or partial knowledge about the properties of quantum systems, including their interaction with the environment, as well as learning ways to better measure and control them.

The methods covered in this Review can be roughly grouped into two conceptually different classes. 
The first consists of methods that are equipped with rigorous performance guarantees, which often consist of pessimistic `worst-case' lower-bounds on the performance. These methods usually suffer from the intrinsic exponential complexity of quantum systems that cannot be avoided in general. The second class consists of heuristics (namely, techniques tailored to specific problems without general performance guarantees) that have been successfully applied to analysing a wide range of systems. 
In the latter case, exponential scaling is commonly avoided by using a specific ansatz that, however, cannot capture a general quantum object. Two important examples of such heuristics are machine-leaning (ML) methods using neural networks (NNs), and Bayesian inference techniques that can incorporate prior knowledge. 
Owing to their high flexibility, these two methods have been widely applied in recent years: for instance, NNs provide a means of building noise resilience into post-processing of measurement data, whereas Bayesian methods allow for the effects of noise to be captured in a probabilistic model of the experiment.

Given a rapidly-expanding relevant literature, a detailed and exhaustive discussion of each technique presented is unfeasible. 
This Review thus illustrates the basic concepts underpinning the main learning techniques and ideas, and points the reader toward relevant references for further insight.
In particular, we combine more detailed explanations for some methods, chosen for their broader applicability, with briefer descriptions of other approaches, referring the reader to landmark publications and existing reviews.
Owing to the intrinsically different nature of heuristic techniques and rigorous-yet-pessimistic approaches, quantitative performance comparisons between these two classes would lead to biased and unjust conclusions.
We therefore refrain from providing a `user manual' for which technique should be used in particular circumstances. 

This Review does not cover approaches based on quantum algorithms, including quantum ML, and we refer the reader to reviews on that subject \cite{Schuld2015, BiamonteNATURE2017, dunjko_machine_2018}. 
Further, there are many excellent reviews that cover and develop aspects related to our discussion, such as randomized-measurement methods~\cite{elben_randomised_measurement_2022}, variational quantum algorithms~\cite{garcia2021learning}, ML techniques for quantum physics \cite{dunjko_machine_2018,carleo_machine_2019,carrasquilla_machine_2020, CarrasquillaPRXQuantum2021,marquardt2021machine,dawid2022modern, krenn_review_2022} and quantum foundations \cite{Bharti_2020}, and the computer-inspired design of quantum experiments \cite{krenn_computer-inspired_2020}. 

\section*{ Learning quantum states}\label{sec_states}

The reconstruction of quantum states from experimental measurements is crucial for characterizing the performance of near-term quantum hardware in terms of fidelity with target states, expectation values of local or nonlocal observables, correlation functions, and other properties. 
Accurately learning quantum states is also a key requirement for studying fundamental physics, such as entanglement properties or the identification of quantum phases of matter. 
In the following, we first discuss the reconstruction of quantum states in a general setting, and then review a variety of assumptions that render this task more efficient. 
Finally, we discuss methods that address cases for which only particular properties of the quantum states are requested, and conclude with an overview on optimal qubit readout.
 
\subsection*{ Quantum-state tomography}\label{section:QST}

The procedure of inferring an unknown quantum state $\rho$ from its measurements is known as quantum-state tomography (QST)~\cite{vogel_determination_1989, raymer_complex_1994, leonhardt_quantum-state_1995,leibfried_experimental_1996,hradil_quantum-state_1997, james_measurement_2001, banaszek_focus_2013, paris_2004}.
Here, we focus on  QST of finite-dimensional discrete-variable systems, such as multiqubit systems.
For tomographic methods in continuous-variable systems, including Wigner tomography to reconstruct the state of an optical mode, we refer to refs.~\cite{paris_2004,lvovsky_continuous-variable_2009}.

A general state of a quantum system corresponds to a positive semidefinite and Hermitian matrix $\rho$ with unit trace~\cite{nielsen_quantum_2011}.
Reconstructing an unknown $\rho$ requires a measurement set that is  
tomographically complete: that is, it must resolve the full space of possible quantum states. In practice, this often requires measurements of $\rho$ in many different measurement settings, and a sufficient repetition of the measurements in each setting, using multiple copies of $\rho$. 
We note that the number of measurement settings can be reduced to one via the use of a polynomial number of ancillary systems 
\cite{d2002universal,banchi2018multiphoton,titchener2018scalable,allahverdyan2004determining}. 
A tomographically incomplete measurement, as well as an imperfect measurement (for example due to noise or an incomplete detector charaterization), generally hinders an accurate estimation of $\rho$.

A common approach to QST is based on maximum-likelihood estimation. This method infers the unknown quantum state as $\rho=\arg\max_{\sigma}P(\boldsymbol{\mu}|\sigma)$, where $P(\boldsymbol{\mu}|\sigma)$ is the probability of the observed measurement results $\boldsymbol{\mu}=(\mu_1,\dots,\mu_m)$ given a quantum state $\sigma$ (refs.~\cite{hradil_quantum-state_1997, banaszek_maximum-likelihood_1999, paris_2004, lvovsky_continuous-variable_2009, shang_superfast_2017}).
The maximization of $P(\boldsymbol{\mu}|\sigma)$ can be performed by different iterative algorithms~\cite{smolin_efficient_2012, rehacek2007diluted, bolduc_projected_2017}. 
However, maximum-likelihood estimation comes with undesirable caveats such as several vanishing eigenvalues of $\rho$ and the difficulty of attributing consistent error bars~\cite{blume-kohout_optimal_2010,christandl_reliable_2012}. These shortcomings have motivated different modern approaches to QST that rely on Bayesian inference, providing a natural way to include prior information (introducing a heuristic element) and to compute error bars~\cite{blume-kohout_optimal_2010,granade_practical_2016,granade_practical_2017,granade_qinfer_2017}, see Box~\ref{sec:bayesian_box} for more details. Note that confidence regions can be estimated by a related approach that is independent of prior assumptions~\cite{christandl_reliable_2012,faist_practical_2016}. 
Finally, in real-time QST~\cite{mahler_adaptive_2013,qi_adaptive_2017,granade_practical_2017}, the measurement data are analysed simultaneously with their recording, and thus  adaptive techniques such as  self-guided QST~\cite{ferrie_self-guided_2014, chapman_experimental_2016, rambach_robust_2021} can be used to optimize the data acquisition procedure. 
Adaptive techniques have also been proposed to reduce the required number of measurements in current quantum computing hardware~\cite{garcia2021learning}, as demonstrated by numerical simulations of variational quantum algorithms that estimate ground states of molecular Hamiltonians.

\onecolumngrid

\begin{mybox}[label = {sec:bayesian_box}]{Bayesian inference}

Bayesian inference is a central method in statistical analysis that prescribes how to update a degree of belief in a hypothesis (for example, that a set of unknown parameters $\boldsymbol{\theta} =(\theta_1, ..., \theta_d)$ has the specific values $\boldsymbol{\phi} =( \phi_1, ..., \phi_d)$) if new observations or evidence (a sequence of measurement results $\boldsymbol{\mu}=(\mu_1, ..., \mu_m )$) are obtained.
It is based on Bayes' theorem, which for parameter estimation takes the form 
\begin{equation} \label{BayesTh}
    P(\boldsymbol{\phi} \vert \boldsymbol{\mu}; c) = \frac{P(\mathbf{\mu} \vert \boldsymbol{\phi}; c) P(\boldsymbol{\phi})}{P(\boldsymbol{\mu}; c)}.
\end{equation}
Equation (\ref{BayesTh}) dictates how to obtain the posterior probability $P(\boldsymbol{\phi} \vert \boldsymbol{\mu}; c)$ from the prior distribution $P(\boldsymbol{\phi})$, using the likelihood $P(\boldsymbol{\mu} \vert \boldsymbol{\phi}; c)$ of observing $\boldsymbol{\mu}$ with measurement settings $c$ if the true phase was $\boldsymbol{\phi}$, and a marginal probability $P(\boldsymbol{\mu}; c)$ for normalization. Note that the choice of an appropriate prior distribution introduces a heuristic component that is often the root of critiques of Bayesian techniques. For a discussion of prior distributions for quantum objects, see refs.~\cite{blume-kohout_optimal_2010,granade_practical_2016,granade_practical_2017}.
In quantum systems, $P(\boldsymbol{\mu} \vert \boldsymbol{\phi}; c)$ is computed using Born's rule.
Estimates and corresponding uncertainties (or confidence intervals) can be obtained directly from the Bayesian posterior: this provides a crucial advantage with respect to the frequentist approach that deduces parameter uncertainties from histograms.

If the measurement results $\boldsymbol{\mu}$ are independent, Bayes' theorem can be also written recursively 
in the form of a Bayesian update when observing the measurement result $\mu_j$, see Fig.~\ref{fig:fig2_bayesian}(a).
Asymptotically in the number of independent measurements $m$ and under mild regularity conditions (such as continuity, regular derivatives and absence of periodicity), the Bayesian posterior converges to a normal distribution
\begin{equation}
    P(\boldsymbol{\phi} \vert \boldsymbol{\mu}) \to \sqrt{\frac{m \det F(\boldsymbol{\theta})}{2 \pi}} e^{-\tfrac{m}{2}(\boldsymbol{\phi} -\boldsymbol{\theta})^TF(\boldsymbol{\theta}) (\boldsymbol{\phi} -\boldsymbol{\theta})},
\end{equation} 
centred at $\boldsymbol{\theta}$ and with covariance given by the inverse of the Fisher Information matrix $F(\boldsymbol{\theta})$
with entries $[F(\boldsymbol{\theta})]_{ij} = \sum_{\boldsymbol{\mu}} P(\boldsymbol{\mu} \vert \boldsymbol{\theta}; c) \partial_{\theta_i} \log P(\boldsymbol{\mu} \vert \boldsymbol{\theta}; c) \partial_{\theta_j} \log P(\boldsymbol{\mu} \vert \boldsymbol{\theta}; c)$, for a given measurement setting. 

A further advantage of Bayesian inference is that it allows a natural integration of optimization techniques~\cite{BerryPRL2000, HigginsNATURE2007, wiebe2016qpe, GebhartPRAPP2021}. 
At each point in time, the current knowledge $P(\boldsymbol{\phi} \vert \boldsymbol{\mu}; c)$ about the parameters can be used to choose optimal experimental settings $c$ for the next measurement. In cases where the optimal measurement settings depend on the (unknown) parameters $\boldsymbol{\theta}$, adaptive techniques can significantly improve the performance and the requirements of the estimation protocol. 
A practical difficulty is that post-processing the Bayesian distribution requires to evaluate it on a grid of $\boldsymbol{\phi}$ values, which can be problematic for broad distributions and in multiparameter scenarios. 
An efficient Bayesian inference for the estimation of multiple parameters~\cite{loredo2004bayesian, granade_practical_2016, GebhartPRAPP2021} thus requires using approximate methods, such as particle-filtering and Sequential Monte Carlo methods~\cite{delmoral2006smc, granade_robust_2012, wiebe2016qpe}, or structured filtering~\cite{granade_2016_structured} when dealing with multiple, equivalent optima.

\end{mybox}
\twocolumngrid

Compressed sensing can be applied when the quantum state $\rho$ to reconstruct is a matrix of low rank~\cite{gross_quantum_2010,flammia_quantum_2012}. In this case, the reconstruction of $\rho$ from the results of random Pauli measurements is reduced to a convex optimization problem, that is, to a semi-definite programme, from which the reconstruction algorithm inherits rigorous performance and convergence guarantees. Compressed sensing can also be used to construct a low-rank estimate that approximates an unknown general state~\cite{riofrio_experimental_2017}. We note that a version of compressed sensing using non-convex optimization was proposed, showing improved performance in accuracy and efficacy under mild regularity assumptions~\cite{kyrillidis2018provable}.

Generally, QST should search through the whole set of degrees of freedom of a quantum state $\rho$, which scales exponentially with the number of qubits $N$. 
For larger $N$, this scaling results in inefficiencies of QST due to the requirement of increasing number of copies of $\rho$ (refs.~~\cite{haah_sample-optimal_2017,odonnell_efficient_2015,kueng_2017_low,chen2022tight}) and the increasing complexity of classical post-processing ~\cite{gross_quantum_2010}. For a review of general complexity requirements of QST, see ref.~\cite{chen2022tight}
For instance, a general tomographic method that can estimate an arbitrary state $\rho$ up to an error $\epsilon$ in trace distance (a common distance measure between density matrices~\cite{nielsen_quantum_2011}) is shown~\cite{haah_sample-optimal_2017,odonnell_efficient_2015} to require at least $\mathcal{O}(4^N/\epsilon^2)$ copies of $\rho$. To reach this optimal scaling, coherent measurements on all copies of $\rho$ have to be performed using a full-scale quantum computer. In the more practical setting of independent (possibly adaptive) measurements of the single copies of $\rho$, optimal strategies (that are related to compressed sensing) have been shown to require $\mathcal{O}(8^N/\epsilon^2)$ copies~\cite{kueng_2017_low,chen2022tight}. 
These obstacles highlight the inefficiency of general QST, already for the current generation of controllable quantum systems~\cite{haffner_scalable_2005}, and the need for new approaches to make QST practical.

\subsection*{ Efficient quantum-state tomography}\label{sec:Efficient_QST}

Several approaches have been proposed to overcome the severe scaling of full QST.
The techniques we discuss below show an improved scalability whenever the quantum state can be described by a specific heuristic ansatz, and they enable QST of highly entangled quantum states of up to about $100$ qubits, which is unfeasible for standard QST methods.

If the state is symmetric under permutations of the qubits, the number of tomographic measurements can be reduced to scale only quadratically with $N$ by using permutationally invariant QST~\cite{toth_permutationally_2010, moroder_permutationally_2012}. 
This method can also be used to estimate the permutationally invariant part of a general state.
Similarly, for systems of identical particles, it is possible to exploit the symmetry of the state to considerably reduce the number of measurement settings~\cite{klose2001measuring,hofmann2004quantum,banchi2018multiphoton,titchener2018scalable}. 
Examples include the reconstruction of a high-spin state \cite{klose2001measuring}, or the reconstruction of the state of many bosons in multiple modes \cite{banchi2018multiphoton}. 

Another method is the use of tensor networks to represent quantum states~\cite{perez-garcia_matrix_2007,cramer_efficient_2010, baumgratz_scalable_2013, lanyon_efficient_2017, wang_scalable_2020}.
The idea is that any pure quantum state can be written as a matrix product state
\begin{equation}
    \ket\psi = \sum_{x_1,\ldots,x_N}{\rm Tr}\left[A_{x_1}^{[1]}A_{x_2}^{[2]}\dots A_{x_N}^{[N]}\right]\ket{x_1x_2\dots x_N},
		\label{eq:mps}
\end{equation}
where $x_n\in\{0,1\}$ labels the computational basis of the $n$th qubit, and $A_{x_n}^{[n]}$ are complex matrices of of dimensions $d_n{\times}~d_{n+1}$, with $d_{N+1}=d_1$.
The largest $d_n$ is called bond dimension, and if it is small, the description of the state and its tomography are efficient~\cite{cramer_efficient_2010}, that is, linear in $N$. 
Even though matrix product states of low bond dimension cannot describe an arbitrary quantum state, they succeed in giving an accurate approximation of ground states of many common Hamiltonians~\cite{cramer_efficient_2010}.
Note that extensions to mixed states and matrix product density operators can be obtained from equation~\eqref{eq:mps} 
with a suitable partial trace~\cite{baumgratz_scalable_2013}.

\begin{figure*}[ht]
\centering
\includegraphics[width=.95\linewidth]{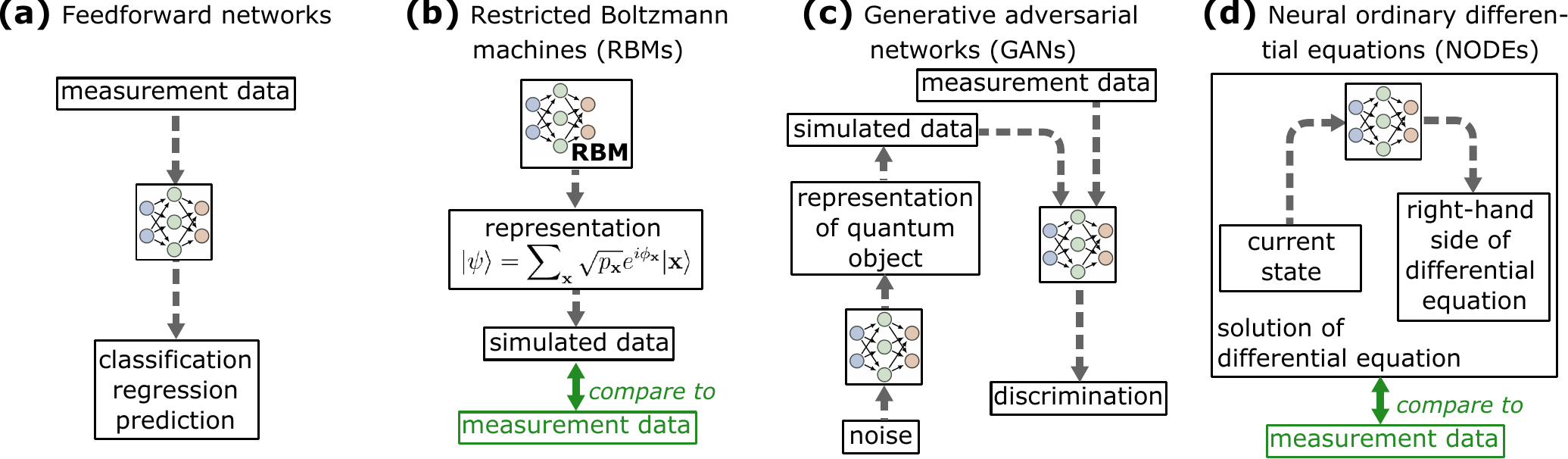}
\caption{\textbf{Different examples of neural networks for learning quantum systems.} Here we consider neural networks (NN) techniques that have been for the post-processing of measurement data. \textbf{a} Feedforward NN for the classification~\cite{carrasquilla_machine_2017,gray_machine-learning_2018,gebhart2021identifying} and regression \cite{goodfellow2016deep} of measurement data, and the prediction of quantum dynamics~\cite{flurin2020using}. \textbf{b} Restricted Boltzmann machines (RBMs) for quantum-state tomography~\cite{torlai_neural-network_2018,carrasquilla_reconstructing_2019}. \textbf{c} Generative adversarial networks for the tomography of quantum states~\cite{ahmed_quantum_2020} and processes~\cite{braccia2021quantum}. \textbf{d} Neural ordinary differential equations (NODEs) for optimal quantum control\cite{schafer2020differentiable,schafer2021control}. Alternatively to NODEs, physics-informed neural networks (that include the differential equation in the cost function) have been used for creating robust quantum gates~\cite{gungordu2020pinngates}.
}
\label{fig:NNs}
\end{figure*}

NNs (Fig.~\ref{fig:NNs}a-d) have been used to perform QST~\cite{CarleoSCIENCE2017, torlai_neural-network_2018,carrasquilla_reconstructing_2019,morawetz_u1_2020,palmieri_experimental_2020}. 
In particular, a wide class of states can be modeled by a so-called restricted Boltzmann machine (RBM) (see Box~\ref{sec_NNbox} and Fig.~\ref{fig:NNs}b), suitably adapted to the quantum setting~\cite{torlai_neural-network_2018, torlai_integrating_2019, melkani_eigenstate_2020}. 
In this method, a general quantum state $\ket\psi = \sum_{\mathbf{x}} \psi(\mathbf{x}) \ket{\mathbf{x}}$ is approximated as $\psi(\mathbf{x}) \propto  \sqrt{p_\mathrm{a}(\mathbf{x})}e^{i\log [p_\phi(\mathbf{x})]/2}$ (up to a normalization factor) where $\mathbf{x}$ labels the computational basis vectors. The parameters 
\begin{equation}
	p_\mathrm{k}(\mathbf{x})=\sum_{\mathbf{h}}e^{\sum_{ij}W^k_{ij}h_ix_j+\sum_{j}b^k_{j}x_j+\sum_{i}c^k_{i}h_i}
\end{equation}
are given by a Gibbs distribution of a NNs consisting of a visible layer ($\mathbf{x}$) and a hidden layer ($\mathbf{h}$), with weights $\{W^k_{ij},b^k_{j},c^k_{i}\}$ to encode the amplitudes ($k=a$) and the phases ($k=\phi$) of $\ket \psi$, and the expressivity of this ansatz is varied by $w$, the number of neurons in the hidden layer $\mathbf{h}=(h_1,\dots,h_w)$.
QST can also be interpreted as a generative adversarial 
game between two players (Box~\ref{sec_NNbox} and Fig.~\ref{fig:NNs}c), a generator trying to produce convincing approximations to the state and a discriminator trying to distinguish real states from generated ones 
\cite{lloyd2018quantum}. As such, classical and quantum generative adversarial networks (GANs) have been used for 
QST~\cite{ahmed_quantum_2020,braccia2021quantum}, requiring fewer measurements when prior knowledge is available. 

\onecolumngrid

\begin{mybox}[label = {sec_NNbox}]{Machine learning and neural networks}

The basic idea of machine learning (ML) is to train a computer to solve a specific task without explicitly instructing it how to operate. Central to this approach is the availability of large amounts of data (or the possibility of synthetically generating them). ML can be used to address various tasks~\cite{goodfellow2016deep} that can be grouped into different types. 
For instance, three important ML tasks are: the classification of data into categories, the regression of functions given their values on data vectors, and the sampling of new data vectors that have a similar distribution to vectors in the given data. 
A central goal of ML algorithms is generalizability -- that is, the computer should succeed in the given task not only for the given training data, but also when new (test) data are provided after the learning phase. 

The basic building block of several modern ML architectures is the artificial neuron. 
These are single-output nonlinear functions $n:\mathbb{R}^n\rightarrow \mathbb{R}$ typically modeled as $n(\mathbf{x})=f(\mathbf{W}\cdot \mathbf{x}+b)$, where $f:\mathbb{R}\rightarrow \mathbb{R}$ is a fixed nonlinear (activation) function, and the weights $\mathbf{W}$ and optional biases $b$ are optimized during the training phase. 
As a single neuron is not sufficient to approximate complex dependencies, multiple neurons are arranged and connected to form a neural network~(NN).
Widely used NN architectures include:
\begin{itemize}
\item  \textbf{Dense feedforward NNs}, with neurons grouped into several layers, where the initial and final layer describe the input and output, respectively. Any neuron can be influenced by any other neuron of the previous layer, and can influence any neuron of the subsequent layer. The intermediate (hidden) layers generate the expressivity of the NN.

\item Feedforward NNs can be equipped with \textbf{convolutional layers}, 
significantly reducing memory and training time requirements. Convolutional layers are particularly suitable to recognize and extract regular patterns. 

\item \textbf{Recurrent neural networks} (RNN), in which subsequent layers can influence previous layers, allowing to retain memory between the processing of different inputs. This approach is often used in time-series or feedback-control scenarios.

\item A restricted Boltzmann machine (RBM), which uses a visible layer that is connected to one or several hidden layers. The interlayer connections represent the weights of a Gibbs distribution, such that they can be trained via Gibbs sampling. RBMs are used to model discrete probability distributions.  

\item Generative adversarial networks (GANs), consisting of two separate and competing NNs, the generator and the discriminator. The discriminator is trained to distinguish genuine data from that generated by the generator. The resulting adversarial game converges when the generator is capable of fooling the discriminator, by generating data that are indistinguishable from genuine ones. 
\end{itemize}
A large enough NN is known to be a universal function approximator \cite{Hornik1993ufa}. However, the size of the NN should be carefully chosen as its trainability can be compromised when chosen too large, and its generalizability may also decrease in presence of a high expressivity and long training schedules (overfitting). 
To train the NN, one must choose a problem-specific cost function that can be minimized via stochastic gradient descent. The way that the NN is trained depends on the given task and is generally divided into three categories:
\begin{itemize}

 \item \textbf{Supervised learning.}  The training data are labeled with their target values: that is, the target function that should be learned by the NN is known for the training data.
 \item \textbf{Unsupervised learning.} The training data are not labelled, and the NN is trained to recognize any type of structure or pattern in the data.
 \item \textbf{Reinforcement learning.} There are no training data, but the NN is connected to an environment. The NN is trained to maximize a reward that is assigned according to the NN's actions.\\\end{itemize} 
\end{mybox}

\twocolumngrid

\subsection*{ Extracting specific features of a quantum state}

If one is only interested in specific properties of $\rho$, the intractability of full QST can be often avoided by tailored measurements and post-processing. 
Most prominently, several methods have been developed to measure entanglement properties~\cite{guhne_entanglement_2009, PezzeRMP2018} such as entanglement entropies (for example using random measurements~\cite{brydges_probing_2019,elben_mixed-state_2020}), metrologically-useful entanglement (such as the Fisher information~\cite{PezzePRL2009, strobel_fisher_2014} or spin-squeezing parameters~\cite{PezzeRMP2018}),  entanglement witnesses (for instance measuring the fidelity to an entangled target state~\cite{lu_experimental_2007}), and entanglement measures (such as the concurrence~\cite{walborn_experimental_2006}). 
NNs have been used to measure the entanglement between two arbitrary subsystems~\cite{gray_machine-learning_2018} and to identify single-mode non-classicality~\cite{gebhart2021identifying} without requiring full QST (Fig.~\ref{fig:NNs}a).

If the measurements of the unknown state $\rho$ are randomly drawn from a fixed (but possibly unknown) distribution, the outcome probabilities of future measurements can be approximately learned (probably approximately correct, or PAC-learned) after merely a linear number of measurements~\cite{aaronson_the_2007,rocchetto_experimental_2019}. This insight has been strengthened in the framework of shadow tomography~\cite{aaronson_shadow_2020, badescu2020}, demonstrating that one can predict the expectation value ${\rm Tr}[\rho O_i]$ of an exponential number of arbitrary observables $O_i$ from the measurements of a polynomial number of copies of $\rho$. However, these methods require a coherent measurement of all copies of $\rho$ in parallel, such that they are out of reach for near-term quantum hardware.
For a review of randomized-measurement techniques, see ref.~\cite{elben_randomised_measurement_2022}.

Finally, the idea of shadow tomography has been adapted to the more practical method of classical shadows~\cite{huang_predicting_2020}  (see also  ref.~\cite{paini2019approximate} for a related approach). As discussed above, these shadows can be used to estimate an exponential number of target observables ${\rm Tr}[\rho O_i]$ from independent measurements of a polynomial number of copies of $\rho$ (coherent measurements of all copies are not required). The price to pay for this simplification is that the scaling also depends on a specific norm of the observables $O_i$, such that for certain observables the scaling becomes exponential . 
To construct the classical shadows, one applies random unitary operators from a fixed ensemble $\Upsilon$, followed by a projective measurement.  
These measurements average to the quantum channel $\mathcal{M}[\rho]=\left|\Upsilon\right|^{-1}\sum_{U\in \Upsilon}\sum_{\mu_U}P_{\mu_U}{\rm Tr}\left[\rho P_{\mu_U}\right]$, where $U$ represents a measurement setting (namely, a unitary transformation applied before a measurement in the computational basis) with possible measurement outcomes $\mu_U$. Each outcome $\mu_U$, described by the projector $P_{\mu_U}$, is observed with probability ${\rm Tr}\left[\rho P_{\mu_U}\right]$. 
For instance, an experimentally accessible measurement ensemble $\Upsilon$ consists of tensor products of single-qubit Pauli measurements~\cite{huang_predicting_2020}. 
We note that if the target functions only include Pauli observables, the formalism can be de-randomised to optimize its performance~\cite{huang_efficient_2021}.
The classical shadow of $\rho$ for the $j$th measurement is then defined as 
\begin{equation}
    \rho_s^{(j)} = \mathcal{M}^{-1}[P_{\mu_j}],
\end{equation}
where $P_{\mu_j}$ is the projector of the $j$th measurement outcome $\mu_j$. 
Because the map $\mathcal{M}^{-1}$ is not a quantum channel, $\rho_s^{(j)}$ is generally not a valid quantum state.
However, each classical shadow results in an estimate ${\rm Tr}[\rho_s^{(j)} O_i]$ of the $i$th target function.
After a statistical inference step (median-of-means), one obtains the final estimates of the target functions. 
Note that classical shadows can also be used to estimate nonlinear functions such as ${\rm Tr}[\rho\otimes \rho O]$~\cite{huang_predicting_2020}. They offer an advantage with respect to maximum-likelihood and Bayesian techniques in certain tasks~\cite{lukens_bayesian_2021}. 
Examples of applications of classical shadows are the measurement of the entanglement entropy~\cite{elben_mixed-state_2020}, or the estimation of the quantum Fisher information~\cite{rath_quantum_2021}. 
Finally, classical shadows combined with classical ML models (trained on previously generated/measured data) were shown to provide provable advantages with respect to classical algorithms that did not use the training data, in tasks such as predicting ground states and classifying topological phases of many-body quantum systems~\cite{huang2021provably}.

\subsection*{ Optimizing qubit readout}

Performing QST of single qubits is greatly aided by the availability of high-fidelity single-shot readout for each qubit state. A single-shot measurement typically consists of a time-resolved signal that is processed to produce histograms relating to the qubit being prepared in the ground or excited state, and a threshold of the readout value is chosen to resolve the two states~\cite{neumann2010single,myerson2008high,vamivakas2010observation,elzerman2004single}. 

Stochastic relaxation of the excited state causes an asymmetry in the readout histograms that decreases the readout fidelity by threshold classification. Clustering methods have been used to identify and discard such relaxation signals and increase fidelity in superconducting qubits~\cite{magesan2015machine}. However, while still increasing overall fidelity, similar methods have produced outlier results of lower quality than thresholding repetitive readouts ~\cite{liu2020repetitive}. Non-linear Bayesian filters have improved threshold fidelities by considering the full time-resolved readout signal while accounting for relaxation~\cite{gambetta2007protocols} and stochastic turn-on times in spin-to-charge conversion readout~\cite{d2014optimal}.
We also note that NNs have improved threshold methods in diamond nitrogen-vacancy (NV) centres which lack single-shot readout at room temperature~\cite{liu2020repetitive}, and NN classifiers trained on synthetic time-resolved data have surpassed the performance of Bayesian filters in quantum-dot spin-qubit readout~\cite{struck2021robust}. Errors in the multiplexed readout of superconducting qubits have been reduced using NNs~\cite{lienhard2022deep} and multiqubit states have been classified using NNs in trapped-ion qubits, where time-binned multi-channel data has provided the highest fidelity~\cite{seif2018machine}. Recurrent Neural networks (RNNs) have been used to predict time-evolving qubit states from experimental noisy measurement traces \cite{flurin2020using}. 

Although the availability of single-shot readout and projective quantum measurements aids learning of quantum systems, it is not necessarily a prerequisite, and averaged readout can be used, in combination for example with Bayesian inference \cite{dinani_bayesian_2019}.

\section*{ Learning quantum dynamics}\label{sec_dynamics}
Reconstructing the dynamics of quantum systems is important for establishing, for instance, channel fidelity in quantum communication, gate fidelity in quantum computing and optimal parameter encoding for sensing applications. 
Following the discussion of QST, we will first review assumption-free methods for learning quantum dynamics and then move to approaches that rely on specific models to simplify the characterization. 

\subsection*{ Quantum process tomography}
\label{sec:qpt}

The task of fully reconstructing the underlying unknown dynamics of a quantum system is called quantum process tomography (QPT)~\cite{Chuang_1997,mohseni_direct_2006}. 
A general quantum process $\Lambda$ that maps a quantum state $\rho$ to a quantum state $\Lambda[\rho]$ is described by a completely-positive trace-preserving map~\cite{nielsen_quantum_2011}.
In the standard setting, one estimates $\Lambda$ by applying the dynamical process to a set of known quantum states $\rho^{\textrm{(in)}}_i$. Each resulting output state $\rho^{\textrm{(out)}}_i=\Lambda[\rho^{\textrm{(in)}}_i]$ is then reconstructed via QST
~\cite{Poyatos_1997, Altepeter_2003, Polino_2020}. For a general (complete) reconstruction of $\Lambda$, the known states $\rho^{\textrm{(in)}}_i$ must represent a basis for all possible initial states, and the measurements of the output states $\rho^{\textrm{(out)}}_i$ should be tomographically complete. 
Therefore, full QPT is even more challenging than QST.
Moreover, enforcing the constraint that $\Lambda$ is a completely-positive trace-preserving map can present further technical difficulties \cite{knee_2018,surawy2021projected}.

Similarly to QST, QPT thus requires a number of measurements and classical post-processing that scales exponentially with the size of the quantum system under study~\cite{mohseni_quantum-process_2008}. Such punishing scaling afflicts, for example,
methods based on maximum-likelihood estimation~\cite{kosut2004optimal}, which additionally show a high sensitivity to errors in initial states, gates and measurements \cite{merkel2013self}.
As in QST, the reconstruction cost can be reduced using compressed-sensing-methods if the process $\Lambda$ does exhibit some latent structure (such as sparsity, low Kraus rank or a simple interaction graph) ~\cite{Shabani_2011,kosut_quantum_2009,Rodionov_2014,kliesch2019guaranteed,seif2021compressed}. Alternatively, quantum processes have been reconstructed using a least-squares estimator (that is not restricted to physical constraints) and then projecting the estimator onto physically-allowed processes~\cite{surawy2021projected}, or by using a gradient-descent optimization of a low-rank Kraus representation of the process~\cite{ahmed2022gradient}.
A second important drawback of standard QPT is the intrinsic self-referential nature of quantum tomography: to calibrate the known initial states, the measurement operators must be known, and to calibrate the measurement operators, the initial states must be known~\cite{nielsen_gate_2020}. This obstacle was addressed with the development of gate set tomography (GST), which reconstructs unknown sets of quantum states $\{\rho_i\}_i$, processes $\{\Lambda_j\}_j$ and measurements $\{E_k\}_k$ (see the next section for more details on the latter) at the same time and in a calibration-free manner~\cite{merkel2013self,blume2013robust,nielsen_gate_2020}. 
Crucially, in this case one reconstructs the objects $\{\rho_i\}_i$, $\{\Lambda_j\}_j$ and $\{E_k\}_k$ by combining them in different circuits (each corresponding to a specific experiment design), resulting in the measurement probabilities
\begin{equation}\label{eq:gst}
   (P_j)_{ik} =  \Tr\left[\Lambda_j[\rho_i]E_k\right].
\end{equation}
GST reflects that, intrinsically, all quantum objects can only be reconstructed up to an equivalence transformation that does not affect the outcomes in equation~\ref{eq:gst}, also called a `gauge freedom', such that the model learned by GST is not unique~\cite{nielsen_gate_2020}. There are several post-processing methods with different levels of sophistication and accuracy performances to reconstruct the quantum objects from the measured probabilities~\cite{nielsen_gate_2020}. To ensure the physicality of the estimated process, the (unconstrained) optimization, for example via gradient descent of the highly non-convex likelihood function, can be alternated with projections onto the space of physical processes~\cite{knee_2018}. Furthermore, in long-sequence GST the tomography can be improved by including measurement data where the processes $\Lambda_l$ are applied several times before measuring the system. GST has been applied in different experimental platforms such as superconducting qubits~\cite{song2019quantum} and trapped ions~\cite{zhang2020error}. In this context, we should mention the calibration-free method of randomized benchmarking~\cite{Knill_2008} which estimates the error rate (or fidelity) of a quantum process from a sequence of random process repetitions
~\cite{Epstein_2014,Rodionov_2014,Granade_2015, Claes_2021}.
This technique, however, does not reconstruct the process matrix $\Lambda$ and thus does not deliver a fully general characterization of the quantum process.
Note also that compressed-sensing methods can be combined with QST \cite{brieger2021compressive} or randomized benchmarking \cite{roth2018recovering} for recovering quantum gates.

Alternative approaches use different heuristic ans{\"a}tze for the quantum process. For example, building on methods introduced for QST, tensor networks 
can lead to greatly reduced resource requirements~\cite{torlai_quantum_2020, gazit_quantum_2019, bennink_quantum_2019}. 
GAN-based approximations of the superoperator $\Lambda$ have also been proposed as a method for QPT~\cite{braccia2021quantum}.
Other methods exploit NNs to generalise QPT to the characterization of time-dependent  spin systems~\cite{Han_2021}, whereas yet another class reconstructs a unitary quantum process by inverting the dynamics using a variational algorithm~\cite{carolan_variational_2020,xue_variational_2021}. RNNs have recently been applied in learning the non-equilibrium dynamics of a many-body quantum system from its nonlinear response under random driving \cite{mohseni2021deep}.

\subsection*{ Reconstructing Hamiltonian quantum dynamics}
\label{sec:gen_learning}

To avoid the general resource-demanding QPT, a commonly-used approximation of the dynamics is given by a heuristic Hamiltonian model for a unitary evolution with a small number of tunable control parameters. Any unitary evolution $U(t,t_0)$ from time $t_0$ to $t$ can be expressed~\cite{breuer2002theory} in terms of a generating Hamiltonian $H$: $U(t,t_0)=\mathcal{T}_\leftarrow \left[ \exp \left(-\frac{2\pi i}{h}\int_{t_0}^t H(t') dt' \right) \right]$, where $\mathcal{T}_\leftarrow$ represents time-ordering, and it simplifies to $U(t,t_0)=\exp(-\frac{2\pi i}{h} H (t-t_0))$ if $H$ is a constant matrix. For a $d$-dimensional system, the problem is then to learn the appropriate $d \times d$-matrix $H(t)$. Leveraging prior information about the system, one can use an appropriate ansatz for $H$ that simplifies the reconstruction.
Nevertheless, reaching an $\epsilon$ average precision in estimating the Hamiltonian parameters, requires a budget in quantum resources (number of measurements and accrued evolution time $t$) that typically scales proportional to $\epsilon^{-2}$~\cite{huang2022scaling}. This can be a limiting factor in devices with limited decoherence times.
A first class of methods relies on obtaining a mathematical model of the dynamical process on the basis of observed time series of measurements ~\cite{di_franco_hamiltonian_2009, cole_PRA_2005, devitt_PRA_2006}. 
Hamiltonian parameters for a few qubits can be retrieved by fitting these time traces or by using Fourier methods. 
More sophisticated approaches rely on using linear systems theory, such as the eigenstate realization algorithm~\cite{zhang_quantum_2014, sone_cappellaro_2017}. In this case, after decomposing the Hamiltonian into Hermitian operators, the evolution can be described by a linear differential equation. 
The algorithm then follows linear systems theory, by adopting a system realization from the experimental data attained at regular time intervals, leading to a transfer function mapping input states to observables, from which one obtains the unknown parameters. More recent versions further refine these techniques by extracting the eigenfrequencies of $H$ from a complex time-domain measurement and then recovering the eigenvectors of the Hamiltonian via constrained manifold optimization~\cite{hangleiter2021precise}.

\begin{figure*}[ht]
\centering
\includegraphics[width=.95\linewidth]{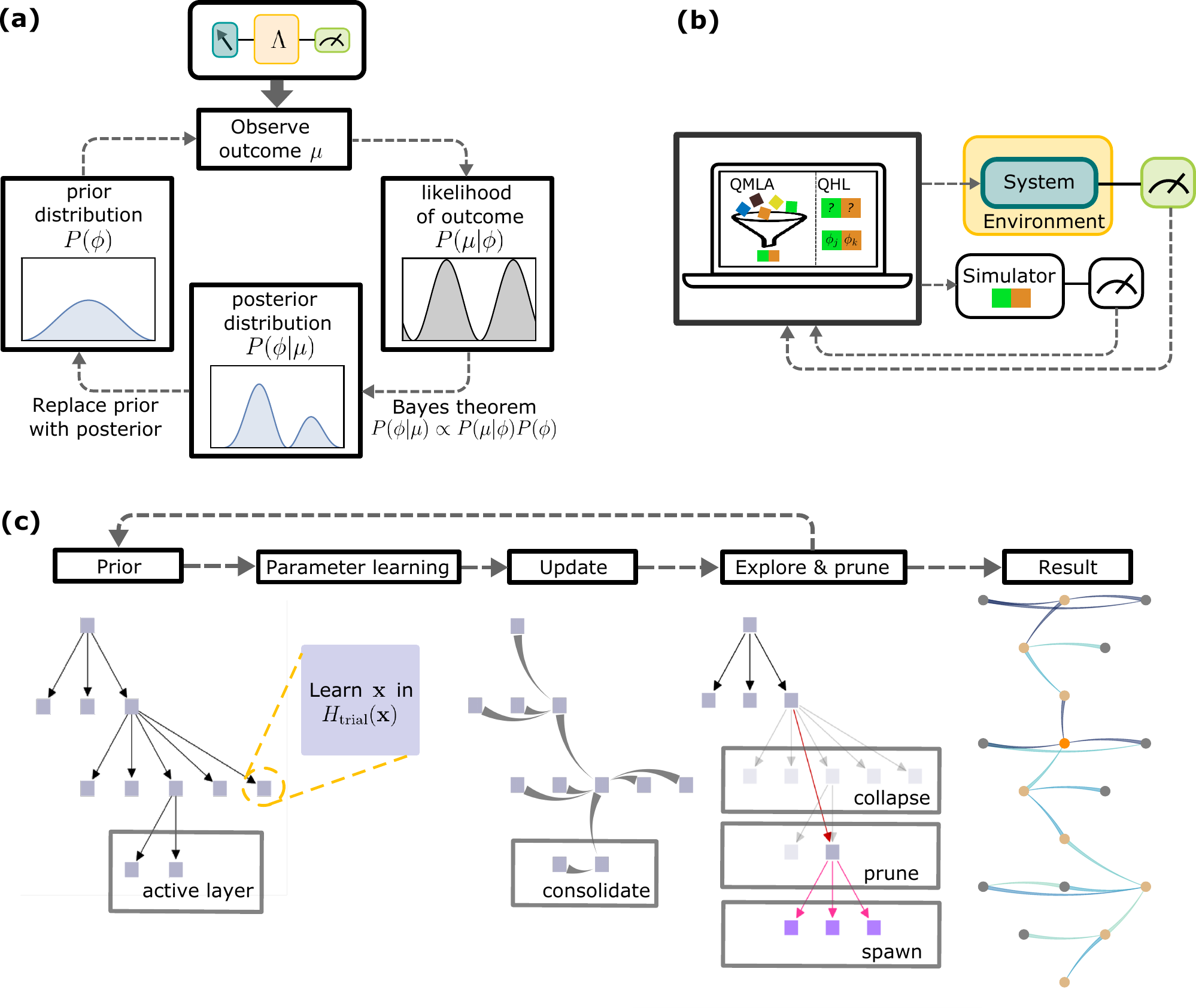}
\caption{ \textbf{Bayesian techniques for quantum model learning.} \textbf{a} Bayesian inference enables to learn the probability distribution for unknown parameters by applying Bayes rules with the measurement outcome of each experiment performed on the quantum system (Box \ref{sec:bayesian_box}). \textbf{b} An overview of the principal subtending protocols for learning quantum dynamics that use a (quantum) simulator~\cite{granade_robust_2012, wiebe_quantum_2014, wiebe_hamiltonian_2014, wiebe_bootstrapping_2015, wang_experimental_2017, hou_2019_11qubit, hincks_statistical_2018, gentile_learning_2021, flynn_2021_qmla}. 
For each iteration $i$, identical known states $\ket{\psi}_{\text{trial}}$ are prepared and fed to both the system to be characterized and the simulator, where $\ket{\psi}_{\text{trial}}$ undergoes evolution for a predetermined time $t$. The latter is assumed to be tunable, to dial the trial Hamiltonian $H_\text{trial} (\bm{x})$, whose parameterized terms are symbolically rendered by coloured squares, which is either known (for example in quantum Hamiltonian learning, QHL) or hypothesized (for example in quantum model learning agent, QMLA). 
The outcome $\mu_i$ collected at the end of each epoch is then fed into the protocol to perform the relevant inference (see also panel b).
\textbf{c} An outline of the protocols that have been demonstrated for QMLA ~\cite{gentile_learning_2021, flynn_2021_qmla}. The information gathered from experiments is stored in a tree structure, whose leaves are candidate Hamiltonians $H_{\text{trial}} (\bm{x})$. The parametrization for those Hamiltonians in the active layer is learned by a method of choice. Trained, active candidate models $(H_{\text{trial}, i}, H_{\text{trial}, j})$ are then pairwise compared according to a metric $B_{ij}$, capturing the relative performance at reproducing the unknown system and stored as edges of the graph. According to the global outcomes, single nodes or entire layers can be discarded, and relevant terms spawned to new candidate Hamiltonians.} 
\label{fig:fig2_bayesian}
\end{figure*}

A second class of methods uses Bayesian inference (see Fig.~\ref{fig:fig2_bayesian}).
For instance, the Bayesian method of quantum Hamiltonian learning (QHL) infers a set of unknown parameters $\mathbf{x}$ for a specific parametrization $H(\mathbf{x},\boldsymbol{\tau})$ (refs.~\cite{granade_robust_2012, wiebe_hamiltonian_2014, wiebe_quantum_2014}) where $\boldsymbol{\tau}$ represents a set of tunable control parameters that typically is just the evolution time $t$. The specific parametrization $H(\mathbf{x},\boldsymbol{\tau})$ offers an efficient reconstruction because the number of parameters $\mathbf{x}$ is typically much smaller than the (exponential) number of parameters necessary to describe the most general $H$.
Building on Bayes' rule (see Box~\ref{sec:bayesian_box} and Fig.~\ref{fig:fig2_bayesian}a), each measurement result $\mu$ (occurring with a likelihood $P (\mu | \mathbf{x},\boldsymbol{\tau})$), is used to update the knowledge about the unknown parameters $\textbf{x}$. 
For few-qubit systems~\cite{granade_robust_2012, granade_qinfer_2017, stenberg_resonant_2014}, $P (\mu | \mathbf{x},\boldsymbol{\tau})$ can be calculated on a classical computer for all necessary combinations of $(\mathbf{x},\boldsymbol{\tau})$.
For larger systems, this classical estimation ultimately becomes intractable~\cite{wiebe_hamiltonian_2014, wiebe_quantum_2014, wiebe_bootstrapping_2015}. QHL thus requires access to a quantum simulator that can implement a controlled evolution by $H(\mathbf{x},\boldsymbol{\tau})$ for different combinations of $(\mathbf{x},\boldsymbol{\tau})$ to experimentally estimate $P (\mu | \mathbf{x},\boldsymbol{\tau})$ (see Fig.~\ref{fig:fig2_bayesian}b). 
QHL has been successfully demonstrated in different experimental platforms and for various tasks, such as characterizing NV centres in diamond~\cite{wang_experimental_2017, hou_2019_11qubit, gentile_learning_2021} or quantum sensing of magnetic fields~\cite{hincks_statistical_2018, santagati_magnetic-field_2019, joas_online_2021}, and can potentially be adapted for quantum control~\cite{wittler_integrated_2021}. 
QHL requires fewer samples and is more robust than, for example, Fourier analysis~\cite{schirmer2015ubiquitous, santagati_magnetic-field_2019}, owing to its Bayesian nature and the use of an adaptive choice of measurement settings according to the cumulative knowledge inferred about the system.

One of the main drawbacks of QHL is that it requires the knowledge of the parametrization $H(\mathbf{x})$. This limitation has been addressed by the use of quantum model learning agents (QMLA) ~\cite{gentile_learning_2021, flynn_2021_qmla} (Fig.~\ref{fig:fig2_bayesian}c), where an artificial agent constructs different candidate Hamiltonian models starting from a number of elementary terms $h_{\nu}$. 
The idea behind is that for example, a large class of $k$-sparse Hamiltonians can be written as a linear combination of tensor products of Pauli matrices acting solely on the $i$-th qubit: $H(\mathbf{x}) = \sum_{i, \alpha} x^i_{\alpha} \sigma^i_{\alpha} + \sum_{i,j,\alpha,\beta} x^{ij}_{\alpha \beta} \sigma^i_{\alpha} \otimes \sigma^j_{\beta} + \ldots $, such that the problem of constructing an appropriate parametrization can be recast in identifying all and only the relevant terms in such an expansion.
Candidate terms $h_{\nu}$ are selected and combined via tree searches or genetic algorithms to generate new candidate models. After a parameter training via QHL, the different candidate terms are systematically compared against remaining candidates using Bayes factors or modified Elo-ratings~\cite{gentile_learning_2021, flynn_2021_qmla}, and the least performing instances are discarded (see Fig.~\ref{fig:fig2_bayesian}c). 

The task of Hamiltonian learning has also been addressed by  deep learning methods. For example, NNs were used to recover Hamiltonians from local measurements of ground states \cite{xin_local-measurement-based_2019}, and RNNs can learn time-dependent target Hamiltonians from time traces of single-qubit measurements without knowledge of the ground states  \cite{che_physrevresearch_2021}. However, the training of NNs is notoriously time-consuming. A possibility to circumvent this problem is to include inductive bias coming from the known physical laws describing the system. Examples of this are `gray-box' approaches ~\cite{peruzzo_QSaT_2020, peruzzo_greybox_2022} that combine a NN as a `black box' describing the Hamiltonian with a physically-understandable `white box' embedding the rules of quantum mechanics, such as state evolution. This approach is able to address uncertainties in the Hamiltonian model, distortions caused by undesired macroscopic dynamics and imperfect measurements.   

Because the space of possible local Hamiltonians has a dimension that scales polynomially with the number of qubits, measuring a polynomial set of local observables (and their correlations) that span the space of local Hamiltonians is sufficient to reconstruct $H$, in contrast with worst-case characterization methods that typically scale exponentially with the size of the quantum system.  
An example of the former is Hamiltonian tomography \cite{wang2015tomo}, where the (polynomial in the system-size) number of unknown parameters in a Hamiltonian which is known to feature solely two-qubit interactions is retrieved tomographically for each two-qubit subset, by applying dynamic decoupling schemes that `isolate' said subsystem while all and only the relevant parameters are being estimated. In this way, system-wide tomography and its expensive scaling can be avoided. Combining these ideas with robust phase estimation methods to replace tomographic measurements, a fast scaling in quantum resources ($\propto \epsilon^{-1}$) was recently proposed \cite{huang2022scaling}. 
Alternatively, if one can prepare a single eigenstate (such as the ground state) of a local Hamiltonian $H$, then $H$ can be reconstructed using only a polynomial amount (in the number of qubits) of measurements and post-processing ~\cite{qi_determining_2019, chertkov2018inverse, greiter2018identify}.
Similar methods can be applied if a generic mixed state that commutes with $H$ can be prepared~\cite{bairey_learning_2019}, and if only local measurements of limited regions of the full systems are available \cite{zhu2019reconstructing}.
These reconstructions eventually reduce to the inversion of a correlation matrix of observables, which, for correlation matrices with vanishing spectral gap, again results in exponential complexity scaling~\cite{evans_scalable_2019}. This obstacle has been overcome in theoretical work~\cite{anshu_sample-efficient_2021, tang_hamiltonian_polynomial}, providing a formal proof that, at finite temperature, a local Hamiltonian $H$ can be reconstructed from only a polynomial number of local measurements on its thermal state $\rho=\exp[-\beta H]/\Tr[ \exp[-\beta H]]$, where $\beta$ is proportional to the inverse temperature. The estimation of the parameters of $H$ is obtained from the estimation of $\rho$, and the latter is reconstructed by maximizing the von Neumann entropy $S(\sigma)=-\Tr[\sigma \log \sigma]$, where $\sigma$ is any state that matches the measurement results. Crucially, it was demonstrated~\cite{anshu_sample-efficient_2021} that the free energy of thermal states is strongly convex with respect to the parameters of $H$, providing rigorous performance and convergence guarantees for the reconstruction.

\subsection*{ Reconstructing open quantum system dynamics}

The Hamiltonian description discussed before is a good approximation of the quantum dynamics if the quantum system is sufficiently isolated from the environment. For a general quantum process, however,  the Hamiltonian description fails and the system has to be treated as an open quantum system\cite{breuer2002theory,leggett1987}. For a memory-less environment, the system dynamics is Markovian and the evolution of the quantum state $\rho$ can be described by the Gorini--Kossakowski--Sudarshan--Lindblad master equation~\cite{gorini1976completely,lindblad1976generators}
\begin{equation}
	 \dot{\rho} = -i[H,\rho] + \sum_j \left(L_j \rho L_j^\dagger - \{L_j^\dagger L_j,\rho\}/2\right),
	\label{eq:lindblad} 
\end{equation} 
where $\{A,B\}=AB+BA$, $H$ is the Hamiltonian, and $L_k$ are the Lindblad operators that describe the dissipative process. The reconstruction of $H$ and $L_k$ from measurement data 
has been dubbed Lindblad tomography~\cite{samach2022lindblad}, where an algorithm based on maximum-likelihood estimation was presented. Alternatively, the generators can be reconstructed from the measurement of time traces and linear system realization theory~\cite{zhang_identification_2015}. Local Markovian dynamics can also be recovered via local measurements when steady states can be prepared \cite{bairey202open}, adopting techniques similar to those outlined for Hamiltonian reconstruction, when knowledge of approximate eigenstates is available. Note that even learning such Markovian dynamics can be significantly more challenging than Hamiltonian learning: recasting equation~\eqref{eq:lindblad} into the equivalent form $\dot{\rho} = \mathcal{L} [\rho]$ shows that the task now involves learning the $d^2 \times d^2$-dimensional Liouvillian superoperator $\mathcal{L}$ for a $d$-dimensional system. However, like in many cases illustrated for Hamiltonian learning, a heuristic approach, and/or prior knowledge regarding expected noise-processes, can permit more efficient parameterizations of $\mathcal{L}$ (see refs.~\cite{cattaneo_PRA_2020, pastori_PRXQuantum_2022}). 

Although often providing an accurate description of the quantum dynamics, the assumptions resulting in a Markovian evolution are not fulfilled in general~\cite{breuer2002theory,rivas2011open, li2018concepts}. 
In non-Markovian dynamics, the state at time $t+\mathrm{d}t$,  $\rho_{t+\mathrm{d}t}$, depends not only on $\rho_{t}$, but also on the system's history at earlier times. 
This evolution can be expressed by the master equation 
$\dot\rho_t = \int_0^t \mathrm{d}s \mathcal K_{t,s}[\rho_s]$ with a time-nonlocal superoperator $\mathcal K_{t,s}$ acting on $\rho_s$~\cite{breuer2002theory}. 
Alternatively, one can use a process tensor formalism~\cite{pollock2018non}, where the time-discretized unitary  evolution operator of the system is interleaved with environmental influence interventions from a `quantum-comb'-like process tensor. 
Although the dimension of this process tensor scales exponentially with the number of time steps, often (for example when memory effects are short-range) it is possible to express the process tensor in compressed matrix operator form (equation~\eqref{eq:mps}), which can be constructed from knowledge of the underlying microscopic model~\cite{jorgensenPT_2019, cygorek2021numerically} or reconstructed from time measurements~\cite{pollock2018non,luchnikov_machine_2020}. Such process tensor tomography generalises QPT, thereby providing access to multi-time correlations, and it has recently been applied to noisy intermediate-scale quantum (NISQ) devices \cite{white2022-PTT}.
Alternatively, heuristic methods based on NNs can be used to reconstruct non-Markovian quantum dynamics 
\cite{banchi2018modelling, krastanov_unboxing_2020}. Generally, these approaches entail reduced physical insight, however, in some cases a degree of interpretability can be preserved, for instance by rendering the master equation \eqref{eq:lindblad} 
non-Markovian through effective time-dependent Lindblad operators that depend on the entire evolution history.
In one such example, the matrix elements of $L_k$ were expressed using RNNs~\cite{banchi2018modelling}, a type of NN capable of modelling long-range memory effects.

\section*{ Learning quantum measurements}\label{sec_measurements}
\label{sec:qmeasure}

The characterization of detectors is commonly given in terms of quantum efficiency, linearity, rate of dark counts, and spectral and temporal response. 
However, these aspects represent only an approximation of the actual operation of a detector and can introduce systematic errors that may strongly affect high-precision measurements or investigations of quantum effects. Quantum detection tomography (QDT)~\cite{LuisPRL1999,FiurasekPRA2001} aims at reconstructing quantum measurement devices without any prior information or approximations.
This characterization plays a key role in any quantum architecture.

Realistic (noisy) detectors are non-projective, and thus
QDT consists in the tomographic reconstruction of a set of positive-operator valued measure (POVM) operators~\cite{LuisPRL1999,FiurasekPRA2001,DArianoPRL2004}. According to the Born rule,
\begin{equation} \label{QDT}
    \Pr(\mu) = {\rm Tr}[\rho E_{\mu}], 
\end{equation}
describes the probability of a generic measurement detection event $\mu$ associated to the POVM operator $E_\mu$, 
satisfying $E_\mu\geq 0$ (which assures $\Pr(\mu) \geq 0$) and $\sum_{\mu=1}^K E_\mu = 1$ (which guarantees $\sum_{\mu=1}^K \Pr(\mu) = 1$).
The standard QDT approach consists of inverting equation~(\ref{QDT}). This uses the experimentally-sampled probability distribution and a suitably chosen set of probe states that span the Hilbert subspace where the POVM elements are defined. 
As a simple example, consider an ideal (von Neumann) projective measurement $\Pr_{\rm id}(\nu) = \langle \nu \vert \rho \vert \nu \rangle$, where $\{ \vert \nu \rangle \}$ is a set of orthogonal states, and model a noisy detection by a stochastic mapping $V_{\mu \nu}\geq 0$ such that $\Pr_V(\mu) = \sum_\nu V_{\mu\nu} \Pr_{\rm id}(\nu)$ is the probability of a detection event~\cite{hetzel_tomography_2022}, depending on the specific choice of $V$. 
As $V_{\mu\nu}$ describes the probability of observing the result $\mu$ when $\nu$ should be ideally observed, it includes the detector's imperfections, and satisfies the normalization condition $\sum_\mu V_{\mu\nu}=1$ for all $\nu$.
In this case, given the observed probabilities $\Pr(\mu)$, QDT corresponds to finding the matrix $V$ that minimises the distinguishability between the probability distributions $\Pr(\mu)$ and $\Pr_V(\mu)$ (for example quantified by the Kullback--Leibler divergence or by the fidelity).  
Such an optimization can be obtained via gradient descent~\cite{hetzel_tomography_2022} or a maximum-likelihood method. 
The corresponding POVM in equation~(\ref{QDT}) reads $E_\mu = \sum_\nu V_{\mu\nu} \vert \nu \rangle \langle \nu \vert$.

Different detector-tomography techniques have been developed and used experimentally, with a main application being the characterization of optical photocounting~\cite{LundeenNATPHYS2009, DAuriaPRL2011, BridaNJP2012} and homodyne~\cite{ZhangNATPHOT2012, GrandiNJP2017} detection, and qubit readout in quantum  computing  machines~\cite{ChenPRA2019}.
With superconducting single photon detectors, QDT has provided a valuable tool to discriminate among different models to explain the basic physics underpinning the detection mechanism \cite{RenemaPRL2014}.
Usual QDT methods require the precise calibration of probe states that, in turn, demands the precise knowledge of the measurement device, thus possibly inducing systematic errors~\cite{MogilevtsevNJP2012, KeithPRA2018} and self-reference~\cite{nielsen_gate_2020}. Apart from the techniques of QST, self-characterizing QDT techniques that do not rely on precisely-calibrated probe states have been demonstrated for single optical qubits~\cite{ZhangPRL2020}, and related ideas have been addressed using NNs~\cite{palmieri_experimental_2020}.

\section*{ Applications in quantum sensing and control}\label{sec_optimization}
In this section, we briefly discuss a few selected examples of applications of learning techniques that rely on classical post and online processing in quantum sensing, imaging and control.

\subsection*{ Quantum sensing and target detection}

Quantum sensors are quantum devices used to estimate a specific quantity of interest, $\theta$, that affects the system's evolution~\cite{DegenRMP2017, PezzeRMP2018, GLM2011, moreau2019imaging,
pirandola2021quantum2}. 
Formally, a quantum sensor is described by an input probe state $\rho$, an evolution $\Lambda_\theta$ and a POVM $E_\mu$. 
The task is to estimate $\theta$ after collecting $m$ measurements (each measurement results being observed with probability $P(\mu|\theta)=\Tr[ E_\mu \Lambda_{\theta}[\rho]]$).
Quantum sensing can thus be seen as a special case of QPT. 
The parameter $\theta$ can be continuous, such as magnetic fields, external forces, accelerations, and so on, or discrete, such as the binary variable certifying the presence or absence of a target, or the pixel value in quantum imaging. 
Quantum systems provide a key advantage in terms of their small size, resulting in high spatial resolution. 
Furthermore, quantum effects such as coherence, squeezing and entanglement can increase the sensitivity of the device~\cite{PezzeRMP2018, GLM2011}. 

When $\theta$ is continuous, 
an important concept is the quantum Cramer--Rao  uncertainty~\cite{helstrom_quantum_1976, BraunsteinPRL1994}
\begin{equation} \label{QFI}
    (\Delta \theta)^2_{\rm QCR} = \frac{1}{m F_Q(\theta)},
\end{equation}
where $F_Q$ is the quantum Fisher Information (QFI).
In a frequentist setting~\cite{LiENTROPY2018}, equation~(\ref{QFI}) provides the smallest possible uncertainty of an unbiased estimate of $\theta$, obtained from the optimization over all possible estimators and POVM~\cite{BraunsteinPRL1994}. In the limit $m\gg 1$, the bound is saturable by the maximum-likelihood estimator.
In a Bayesian setting (see Box \ref{sec:bayesian_box}), equation~(\ref{QFI}) provides the asymptotic ($m \gg 1$) posterior variance using an optimal POVM.
NNs have been trained to construct estimators~\cite{cimini_calibration_2019, cimini_calibration_2021, NolanQST2021} or Bayesian posterior distributions~\cite{nolan_machine_2020} based on supervised learning techniques and using limited calibration data.
In both cases, a well-calibrated network can be used for post-processing data that saturate equation~(\ref{QFI}) for a sufficiently large number of measurements.
These NN-based estimation methods are especially valuable when one is lacking simple models for the output probability distribution of the sensor, for example when the number of possible measurement events is large.

A notable result in quantum sensing is the relation between uncertainty and entanglement~\cite{PezzeRMP2018, TothJPA2014}: separable states of $N$ qubits that undergo a collective spin rotation (as common in many applications) can achieve, at best~\cite{PezzePRL2009}, a QFI $F_Q=N$.
Entanglement is necessary~\cite{PezzePRL2009} for $F_Q>N$, whereas genuine multipartite entanglement is necessary~\cite{HyllusPRA2012, TothPRA2012} to achieve the ultimate limit $F_Q=N^2$. 
Interestingly, multipartite entanglement witnessed by the QFI~\cite{HyllusPRA2012, TothPRA2012} is found in a variety of many-body phenomena such as quantum phase transitions, quantum chaos, quenches, random states, and so on.

In the case of the joint estimation of $d$ independent parameters $\theta_1, ..., \theta_d$, the $d\times d$ covariance matrix of estimators is bounded by the inverse of the QFI matrix \cite{helstrom_quantum_1976, AlbarelliPLA2020}.
In general, when estimating multiple parameters, there is no optimal observable with which the quantum Cramer--Rao bound can be saturated. 
This problem is linked to non-commutativity of the optimal measurements for the different parameters~\cite{helstrom_quantum_1976, AlbarelliPLA2020} and the more involved Holevo--Cramer--Rao bound must be used instead \cite{RafalPLA2020}. 
For commuting encoding transformations, the interplay between mode and particle entanglement and the QFI matrix has been discussed in ref. \cite{GessnerPRL2018} and experimentally investigated with optical qubits \cite{LiuNATPHOT2020}.  

When the parameters are discrete, quantum advantage cannot be quantified via the quantum Cramer--Rao bound of equation~(\ref{QFI}), and alternative strategies have been proposed~\cite{hassani2017digital}. A common heuristic approach is first to define a problem-dependent cost function and then compare the performance of entangled sensors with strategies based on classical sources. 
Entangled sensors have been used to define a quantum support vector machine able to more accurately classify the presence of a target \cite{zhuang2019physical}. Moreover, entangled probes can be used to read binary images with enhanced precision and perform more accurate image classification \cite{banchi2020quantum}.
Although for full imaging each pixel must be accurately reconstructed, for pattern recognition errors are tolerated and less precise detectors are allowed. This intuitive idea was formalized in ref. \cite{banchi2020quantum}  by showing that the error with optimal detectors decreases exponentially with the number of bits that must be flipped to switch class. 
Finally, quantum imaging can be enhanced by deep learning techniques \cite{li2021fast,picard2019deep,harney2021ultimate}, in order to increase the fidelity, discover deeper structure in data and overcome errors due to shot noise and background noise.

\subsection*{ Adaptive methods for quantum sensing}

\begin{figure*}[ht]
\centering
\includegraphics[width=.95\linewidth]{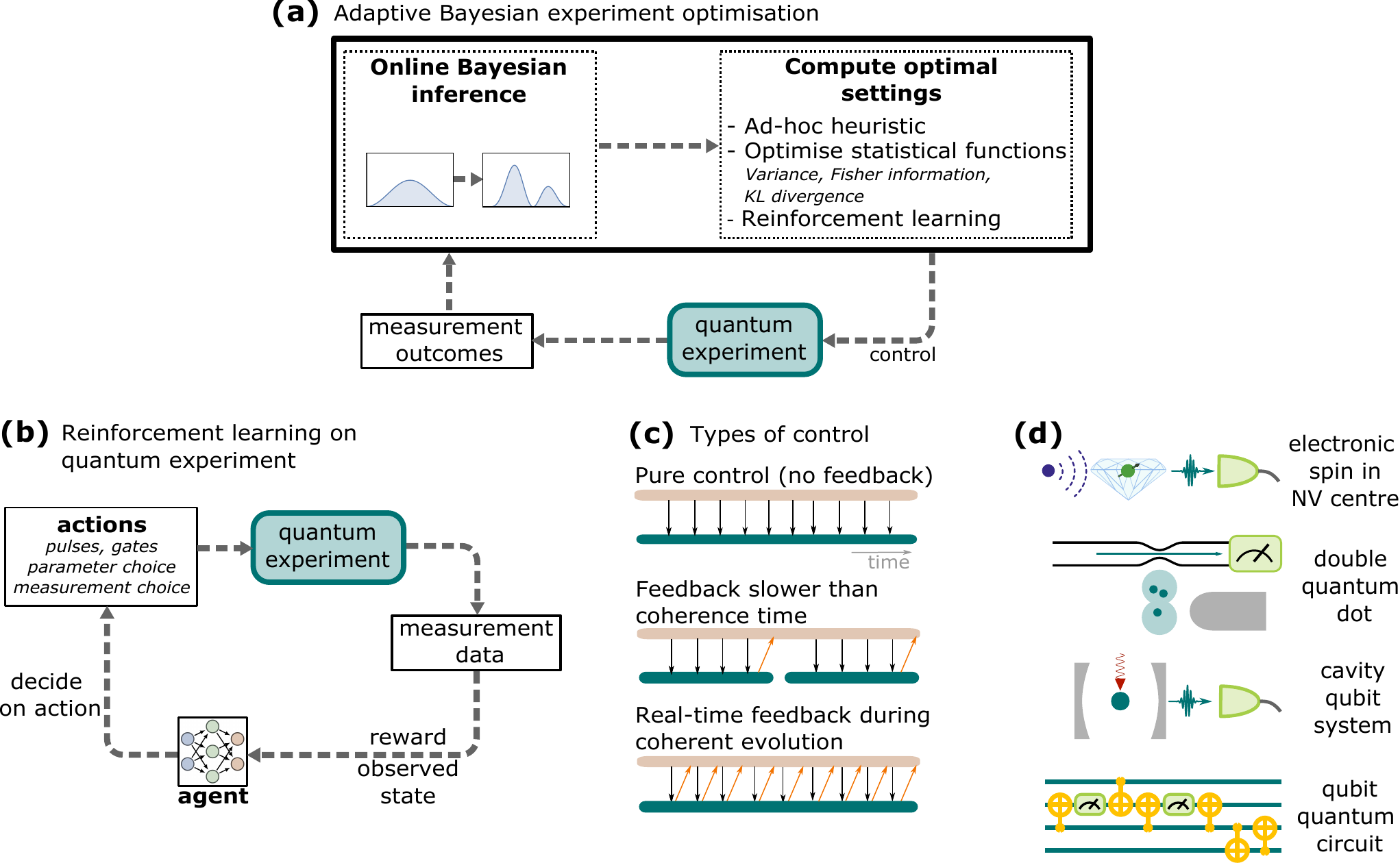}
\caption{
{\bf Optimizing quantum experiments.} {\bf a} Example of adaptive experimental control, through Bayesian inference. After each measurement, the outcome is used to update the current probability distribution for the parameters of interest. This probability distribution is then used to compute optimal settings for the next measurement, through ad-hoc heuristics, the optimization of statistical quantities (variance, Fisher information, Kullback--Leibler (KL) divergence, and so on). 
{\bf b} Schematic feedback loop optimized by reinforcement learning (RL). Measurement results are fed into an agent that decides on the next action to apply to the quantum experiment. {\bf c} Different strategies for optimized quantum control; time evolution, with control pulses as black arrows, measurements as orange arrows, agent in brown, and the quantum experiments' coherent evolution intervals in green. {\bf d} Adaptive Bayesian inference and RL have been applied to or suggested for several experimental quantum systems (schematically shown in the panel), such as NV centres ~\cite{bonato_optimized_2016, santagati_magnetic-field_2019, dushenko_sequential_prappl2020, joas_online_2021, mcmichael_sequential_2021, childress2022}, quantum dots \cite{lennon2019efficiently,nguyen2021deep}, cavity qubit systems \cite{sivak2022model,porotti2021deep} and multiqubit systems with gates applied as actions and subject to projective measurements \cite{fosel_reinforcement_2018}. 
 }. 

\label{fig:fig4_optimization}
\end{figure*}

Quantum sensors can be optimized by harnessing adaptive protocols. Such protocols use newly acquired data to compute the optimal experimental settings from the current knowledge about unknown parameters being estimated.
Although the optimal setting of the sensor (for instance the optimal POVM satisfying equation~(\ref{QFI})) generally depends on the unknown parameters, adaptive strategies can converge to the optimal configuration with an increasing number of measurements.
Bayesian methods (see Box \ref{sec:bayesian_box}) are well suited for adaptive optimization~\cite{HigginsNATURE2007, FidererPRXQuantum2021} (Fig.~\ref{fig:fig4_optimization}a).
Several adaptive protocols have focused on the optimization of measurements \cite{BerryPRL2000, HigginsNATURE2007}, including numerical techniques such as particle swarm~\cite{ hentschel_efficient_2011} and differential evolution algorithms~\cite{lovett_differential_2013}.
Reinforcement Learning (RL) has been used to discover the whole sequence of optimal adaptive measurements of a qubit in the Bayesian framework \cite{FidererPRXQuantum2021}, and to devise feedback control strategies for frequency estimation for a bosonic field \cite{FallaniPRX2022}. 
The optimization of probe states, for example in order to extend the bandwidth of entanglement-enhanced sensitivity, has been considered using analytical methods \cite{PezzePRL2020, PezzePRXQuantum2021}, as well as 
variational~\cite{KaubrueggerPRL2019, KaubrueggerPRX2021} and NN strategies~\cite{HainePRL2020}.
Besides the optimization of measurements and probe states, another possibility is to modify parameter encoding by adding control Hamiltonians \cite{ pang_optimal_2017,xu_generalizable_2019,SchuffNJP2020, XiaoNPJQI2022}, optimized via RL.
These techniques become particularly useful in a sensor network due to the increasing complexity of the system and the multi-dimensional parameter space \cite{CiminiARXIV2022}.

Experimentally, adaptive techniques have been used, for example, to improve the sensitivity of optical phase sensing~\cite{HigginsNATURE2007, DaryanooshNATCOMM2018, valeri_experimental_2020}, or in quantum sensors based on the single spin associated with the NV centre in diamond, a system widely used for nanoscale magnetic mapping and magnetic resonance~\cite{bonato_optimized_2016, santagati_magnetic-field_2019, dushenko_sequential_prappl2020, joas_online_2021, mcmichael_sequential_2021, childress2022}.
Owing to the time constraints, a key consideration for online sensing is the cost of the processing procedure: simplified near-optimal methods might perform better than optimal but computationally intensive ones. This is especially important in the tracking of time-dependent magnetic fields in d.c. magnetometry ~\cite{bonato_tracking_2017, hincks_statistical_2018, santagati_magnetic-field_2019}.

\subsection*{ Learning control strategies for quantum systems}
Reconstructing the dynamics of a quantum system presents opportunities to establish a degree of control over the environment, for example with the goal of reducing the decoherence of the system. QHL has been used to suppress decoherence for a single spin qubit by a real-time compenzation of nuclear-spin bath fluctuations~\cite{shulman_suppressing_2014} and classical noise~\cite{mavadia_prediction_2017}. 
If the environment can be described by a quantum system, the learning process itself can be considered a control tool because the back-action of a quantum measurement perturbs the quantum state, for example by projecting the system to an eigenstate of the measurement~\cite{cappellaro_spin-bath_2012}. 
Even though the random measurement outcomes result in random quantum states of the system, feedback can be used to make the control system deterministic~\cite{blok_manipulating_2014}, following the idea of quantum error correction~\cite{nielsen_quantum_2011}. Protocols have been suggested to extend a qubit's coherence by sequential Hamiltonian learning with real-time feedback from the measurement outcomes~\cite{Scerri_2020}. 

Other applications of ML methods in quantum technologies include the design of quantum gates~\cite{Innocenti_2020, Gao2020} or the identification and design of circuits for quantum error correction~\cite{valenti_hamiltonian_2019}.
Open-loop optimal control allows some improvement of quantum gates, but is limited by the quality of the model available for the system and the control sequences. An alternative framework has been proposed \cite{wittler_integrated_2021}, integrating  calibration, characterization and control in a single process. In this approach, one alternates -- in a closed adaptive loop --  between learning the dynamical model that best explains the system dynamics from the experimental data, and designing better control strategies of the system, based on the learnt model. Using adaptive techniques, a model-based simulation is used to create control schemes, the result of which is then compared with the experimental observation and optimized~\cite{machnes_2022}.

Model-free RL is another interesting general approach to learn control strategies, where an agent (often implemented via a NN) observes and controls an `environment', which might be a quantum device (see Fig.~\ref{fig:fig4_optimization}b). The dynamics of the environment need not be known in advance: the agent will implicitly learn an approximate model while trying to find a good feedback-control strategy. Overall, the agent tries to optimize a reward (defined in terms of final state fidelity for the case of state preparation). In so-called policy gradient approaches, this is done by implementing a parametrised stochastic policy $\pi_\theta(a|s)$ -- the probability of an action $a$ given an observed state $s$ -- and changing by gradient ascent the probabilities in any action sequence to maximise the average cumulative reward (see elsewhere~\cite{marquardt2021machine} for more details).
In theoretical studies, RL has been used to discover feedback strategies for quantum error correction  \cite{fosel_reinforcement_2018}, to obtain qubit control pulses for state preparation \cite{bukov2018reinforcement,niu2019universal}, and to solve many other tasks (see Fig.~\ref{fig:fig4_optimization}c,d). It is beginning to be implemented on experimental platforms, for example for superconducting qubit gate synthesis \cite{baum2021experimental} and also for adaptive characterization of quantum systems \cite{nguyen2021deep}. A particular challenge consists in extracting the reward reliably from available experimental measurements and to make sure that the agent can be trained in a truly model-free way without any assumptions \cite{sivak2022model}.

\section*{ Outlook}

In this Review, we have discussed different techniques that rely on classical post-processing and adaptive optimization to learn information about quantum systems.
In particular, owing to the exponential complexity of quantum states and dynamics and the intrinsic probabilistic nature of quantum measurements, efficient learning techniques are essential for tasks such as the reconstruction, validation and optimal control of quantum systems.
The fast progress of this field take advantage of efficient numerical methods already developed in computer science and statistics to solve demanding problems in quantum physics.
In the following, we discuss a few promising perspectives and possible future developments. 

First, very few detailed comparisons~\cite{schwemmer_experimental_2014,huang_predicting_2020,lukens_bayesian_2021,merkel2013self,FidererPRXQuantum2021,LiENTROPY2018, valenti_PRA_2022} of the different learning methods we discussed in this Review exist in the literature, making it hard to give prescriptions for which method one should use for a task under given assumptions. Only a small minority of the techniques are supported by a rigorous complexity analysis, and most are based on heuristics. We believe the quantum technology and ML communities should address these issues, working on both provable complexity analyses of heuristic methods and numerical comparisons of different techniques for specific problems. 
Results providing rigorous performance (and advantage) guarantees for NN approaches (that are usually based on heuristics) for several tasks~\cite{huang2021provably} represent an important first step in this direction.

Second, despite often being treated separately in the literature, the reconstruction of quantum states, dynamics, and measurements must be approached as a whole~\cite{nielsen_gate_2020}. This problem is addressed, for example, by the techniques of self-calibrating quantum tomography~\cite{MogilevtsevNJP2012,KeithPRA2018,ZhangPRL2020,palmieri_experimental_2020} or GST~\cite{merkel2013self,blume2013robust,nielsen_gate_2020}. A further example is closed integrated learning loops~\cite{wittler_integrated_2021}, which are finding more and more experimental applications. In this case, the information obtained from Hamiltonian learning techniques is used to optimally design control gates and pulses, which, in turn, are used to further refine the characterization of the system. Another possible road to calibration-free methods is based on NNs. Instead of finding and calibrating an explicit model for the system, NNs can be used as `black boxes' to model quantum systems. For instance, RL techniques have been suggested to generate `system-agnostic' heuristic methods for sophisticated adaptive parameter estimation~\cite{FidererPRXQuantum2021}. We expect that NN-based techniques and RL will be increasingly adopted by experimentalists, removing the need for a manual derivation of experimental models and design. Additionally, RL generally presents a new approach for discovering from-scratch feedback-control strategies for quantum devices, for example for state and device characterization or quantum error correction. Although RL works successfully for several tasks in simulations and also in first experiments, there are still multiple demanding challenges to be solved in this domain. Among them are the need for the network agent to find both a good control strategy and an interpretation of the noisy measurement data, the technical hurdles in experimentally implementing real-time feedback using NNs, and the construction of good reward functions that help guide the optimization while being accessible from experimental data.

In the context of ML-methods, one open problem is the lack of physical interpretability of `black-box' NN-approaches. One promising route to circumvent this issue is to incorporate knowledge from the known physical laws describing the system in the NN-description. One option consists of neural ordinary differential equations (NODEs), see Fig.~\ref{fig:NNs}d: these include the system's differential equations ~\cite{chen2018advances} or the system's Lagrangian function~\cite{cranmer2020lagrangian} into the structure of the NN. NODEs have been applied, for example, to the optimal control of a qubit
~\cite{schafer2020differentiable,schafer2021control}. A further technique is physics-informed neural networks that model the solution of the system's differential equations by directly including the latter in the cost function of the NN~\cite{Raissi2019pinn}.
Early applications of physics-informed neural networks explore simple instances of the Schr{\"o}dinger equation~\cite{Raissi2019pinn, bu2020compet_pinn} and the possibility of controlling dynamically-corrected quantum gates~\cite{gungordu2020pinngates}.

A final promising future perspective comes from the recent developments of quantum computing hardware. Whereas a classical computer processes the outcomes of quantum measurements, a quantum computer can directly handle `raw' quantum states from an experiment. The main advantage of this approach is that a quantum computer can perform joint measurements on several qubits, exploiting quantum coherence and entanglement. For instance, it has recently been shown that joint quantum measurements lead to substantial advantages for learning quantum systems~\cite{google_science_2022}, as they outperform single measurements in distinguishing quantum states even when the states are uncorrelated. An obstacle for learning quantum states with a quantum processor is the need for a quantum-coherent interface between the quantum system under study and the quantum processor. This will require further research in quantum architectures where quantum sensors, devices and systems are interfaced to a quantum computer preserving quantum coherence. The fact that advantages in learning quantum systems can already be demonstrated by current noisy intermediate-scale quantum devices~\cite{AruteNATURE2019,ZhongSCIENCE2020}, opens up exciting possibilities that will become more and more important as we advance towards full fault-tolerant quantum computing. 

\section*{Acknowledgements}
The authors thank Chris Ferrie for useful discussions. CB is supported by the Engineering and Physical Sciences Research Council (EP/S000550/1 and EP/V053779/1), the Leverhulme Trust (RPG-2019-388) and the European Commission (QuanTELCO, grant agreement No 862721). NA acknowledges support by the Royal Society (URF/R1/191150), EPSRC Platform Grant (EP/R029229/1), the European Research Council (grant agreement 948932) and FQXi Grant Number FQXI-IAF19-01.
LB's work is supported by the
U.S. Department of Energy, Office of Science, National
Quantum Information Science Research Centers, Superconducting Quantum Materials and Systems Center
(SQMS) under the contract No. DE-AC02-07CH11359, and by the INFN via the QubIT, SFT and INFN-ML projects. 
VG and LP acknowledge financial support from the European Union’s Horizon 2020 research and innovation programme—Qombs Project, FET Flagship on Quantum Technologies Grant No. 820419.

\end{document}